# Efficiency, Robustness and Stochasticity of Gene Regulatory Networks in Systems Biology: λ Switch as a Working Example

Xiaomei Zhu[1], Lan Yin[2], Leroy Hood[3], David Galas[3], and Ping Ao[4],*

[1] GenMath, Corp. 5525 27th Ave.N.E., Seattle, WA 98105, USA
[2] School of Physics, Peking University, Beijing 100871, PR China
[3] Institute for Systems Biology, 1441 N. 34 St., Seattle, WA 98103, USA
[4] Department of Mechanical Engineering, University of Washington, Seattle, WA 98195, USA

**November 30 (2005); updated February, 7 (2006)**

## Summary

Phage λ is one of the most studied biological models in modern molecular biology. Over the past 50 years quantitative experimental knowledge on this biological model has been accumulated at all levels: physics, chemistry, genomics, proteomics, functions, and more. All its components have been known to a great detail. The theoretical task has been to integrate its components to make the organism working quantitatively in a harmonic manner. This would test our biological understanding and would lay a solid fundamental for further explorations and applications, an obvious goal of systems biology. One of the outstanding challenges in doing so has been the so-called stability puzzle of λ switch: the biologically observed robustness and the difficulty in mathematical reconstruction based on known experimental values. In this chapter we review the recent theoretical and experimental efforts on tackling this problem. An emphasis is put on the minimum quantitative modeling where a successful numerical agreement between experiments and modeling has been achieved. A novel method tentatively named stochastic dynamical structure analysis emerged from such study is also discussed within a broad modeling perspective.

**Running title:** systems biology study of λ switch

**Keywords:** phage λ, genetic switch, robustness, efficiency, cooperation, stochastic processes, dynamical landscape, systems biology

* Corresponding author. E-mail: aoping@u.washington.edu; fax: (206) 685-8047; ph: (206) 543-7837

*"I assign more value to discovering a fact, even about the minute thing, than to lengthy disputations on the Grand Questions that fail to lead to true understanding whatever."*

Galileo Galilei (1564-1642)

## 1. Introduction

The completion of Human Genome Project prompts the biological and medical research into a new phase never experienced in biology previously. It is evident that a vast uncharted territory is luring ahead with tremendous promises [1]. Great questions with important conceptual and practical implications have been asked and discussed [2-4]. Speculations on the general principles underlying those great questions and general methodologies to solve them have been extensively debated since the beginning of this century [5,6]. One of the present authors has been steadily promoting such exposition and has been contributing to this trend [7]. Such efforts are needed not only as "a call to arms", they also help to define the various emerging fields. Nevertheless, in the practical research a full range of endeavors has to be explored. New tools will be invented to solve new problems and to try on remaining old problems. In the present review we therefore turn our attention to the other side of consideration, not as an indication to underestimate the value of grand themes but as an example to balance the grandeur. Instead of asking general questions and receiving limited answers, we wish to ask limited questions on a limited system and to find as complete answers as possible, along with a few general answers. We have been carrying this effort during past few years. Such a methodology has been very effective ever since the dawn of modern science, first exemplified by Galileo. Specifically, we will focus our attention to the robustness and stability of a genetic switch [8,9] in phage λ, arguably the biological model started the modern molecular biology [10].

In modern information age switch-like structure is a building block in all architectures. It is the realization of the binary digit, the unit of information and the "atom" today. As biology has been increasingly viewed as an information science [11-13], it would be desirable to have a thorough understanding of this building block. Indeed, detailed analyses have demonstrated that the response of a complexity network is often dealt with various switches [14] and that genetic networks were shown to have the computational ability [15]. By drawing a close analogy to the integration circuitry in electronic wiring board, this methodology has been successfully employed in the modeling of genetic regulation during the earlier developmental stages in sea urchins [16]. Nowadays, the studying of switching in biology has been ranging from responses to environmental changes [17,18], developmental biology [19-21], neural networks [22,23], physiological response [24,25], genetic regulation [26-29], signal transductions [30], memory effect [31,32], olfactory perception [33], synthetic biology [34], biotechnological applications [35-38], to photosynthesis [39] and many other areas [40-43]. Even in cell cycle processes, if viewing such a process not as driving by a cycling engine but as what controlled by a traffic light, the switch-like structure is likely to play a dominant role [44-46]. Switch has indeed established itself as one of the fundamental elements in biological processes and as a paradigm for both experimental and theoretical studies in biology.

Why then has so much effort been expended on studying on a particular virus genetic switch, the λ switch? To paraphrase what Ptashne already stated [8], this is a fair question desiring a clarification at beginning. After all, every case in biology is at least partly accidental and special, and workings of every organism have been determined by its evolutionary history, and the precise description we give of a process in one organism will probably not apply in detail to another. Thus both robustness and stochasticity in bio-structure must be included and carefully studied. This has been well illustrated in the context of the fundamental biological processes such as mentioned in previous paragraph. As already indicated above, at various stages of development, depending in part on environmental signals, cells choose to use one or another set of genes, and thereby to proceed along one or another developmental pathway. It would be of great value to know what molecular mechanisms determine these choices. Hence, the λ life cycle is indeed a prototype for this problem with the structure of feedback loops and the effect of stochasticity. In addition, we have a nearly complete understanding of all its parts, its genome was in fact known [47] long before the completion of Human Genome Project, and the corresponding quantitative knowledge has been accumulated at all levels: physics, chemistry, DNA, protein, and functions [8,48,49]. Despite such a long history of quantitative studies, the stability and robustness of the λ switch remained as one of outstanding puzzles for computational biology at least till 2004 [9,50,51]. The theoretical challenge has been to put all its components together as a harmonic working organism, one of major tasks of systems biology.

In addition, one might wonder the value of using a quantitative and detailed modeling. Biological theories are generally known for their descriptive nature. For example, when Darwin presented his evolutionary theory, no single equation had been used. It was rather remarkable that though one of Darwin's main predictions, the age of Earth, was in direct conflict with known physics at his time, it was physics not Darwin's theory that was later gone through a fundamental transformation to resolve this glare contradiction, to the good of both physics and biology. Nevertheless, it would be wrong to conclude that a quantitative method would be of no use in biology. In fact, some subfields in biology, such as physiology and population genetics, are among those most mathematical in natural sciences [3]. As biology is becoming an information science [11-13], more would be so in the future. The important question is that what would be the right framework of mathematical description [9,52]. It is true that an excessive use of mathematical language, which might be attractive to a modeler, generally does not enhance the understanding of a specific biological phenomenon [53]. For example, with excessive parameters any phenomenon can be described by a set of equations. Such a situation is not acceptable under Ockham's razor. The other extreme is to look for effective description with the hope to capture the biological essence. This later description is necessarily gross and qualitative, though extremely popular and particularly successful in biology. However, many features are obviously left behind by such an approach. It would be desirable to have a detailed quantitative study which can bridge those two types of approaches. The phage λ genetic switch provides precisely one of the excellent opportunities in biology to do so [54]: One side is an on-off Boolean type description for the genetic switch and other side is the detailed physical and chemical equations.

The rest of the review is organized as follows. Salient biological experimental studies on phage λ switch are summarized in section 2. Its key biochemical modeling elements are summarized in section 3. The stochastic dynamical structure analysis method is discussed in section 4 within the minimal quantitative model of phage λ. Calculated results and the comparison to biological data are discussed in section 5. In section 6 we summarize what have been done and place the minimum quantitative modeling methodology in a broader context. In section 7 the research effort on λ switch is put into an optimistic outlook.

## 2. Phage λ Genetic Switch

### 2.1. Phage λ life cycle

Bacterophage λ is a virus that grows on a bacterium [8,55,56]. It is one of the simplest living organisms. Almost all its parts have known during past 50 years. The genome of phage λ consists of a single DNA molecule wrapped in a protein coat. Upon infection of the host *E. coli* cell, the phage λ injects its genome inside the bacterium and leaves the protein coat outside. Inside the bacterium it chooses one of two modes of growth. Phage λ uses molecular-genetic apparatus of the cell for running and executing its own ontogenetic subprograms to produce new λ phage particles, resulting in the lysis of the cell. Or, it can establish dormant residency in the lysogenic state, integrating its genome into the DNA of its host and replicate as a part of the host genome. In these two different life cycles, different sets of phage genes are expressed as a result of molecular interactions. Realistic modeling of the robustness and stability of such a process has been remained one of the most important challenges in biocomputation and bioinformatics.

Because the analogous molecular interactions to phage λ are likely to underlie many developmental [8] and epigenetic [56] processes, one wishes to acquire deep understandings on the regulation of major biologic functions on molecular level through the study of the genetic switch of phage λ. One of these functions is the programming of the epigenetic states: The ways the phage decides if it is going to follow lysogenic or lytic growing state. Over past five decades, the extensive biological investigations have provided a fairly good qualitative picture in this respect. There exists a plausible scenario to guide the understanding of the experimental observations [8].

The maintenance and operation of the genetic switch is another function performed by the gene regulatory network [8,55,56]. The phage growing in lysogenic state remains latent unless it is provoked. For example, switching to lytic state happens when a signal is sent to activate RecA proteins which cleave CI monomer, sending the phage into lytic growth. It is observed that the phage λ genetic switch is both highly stable and highly efficient. When the phage grows in lysogenic state, it remains latent for many generations. Spontaneous induction happens less than once in a million cell divisions. Once the phage is exposed to an appropriate signal, it changes to lytic state at almost 100% rate. Such a coexistence of stability and efficiency of the genetic switch in phage λ has been considered a mystery from the theoretical and mathematical modeling side.

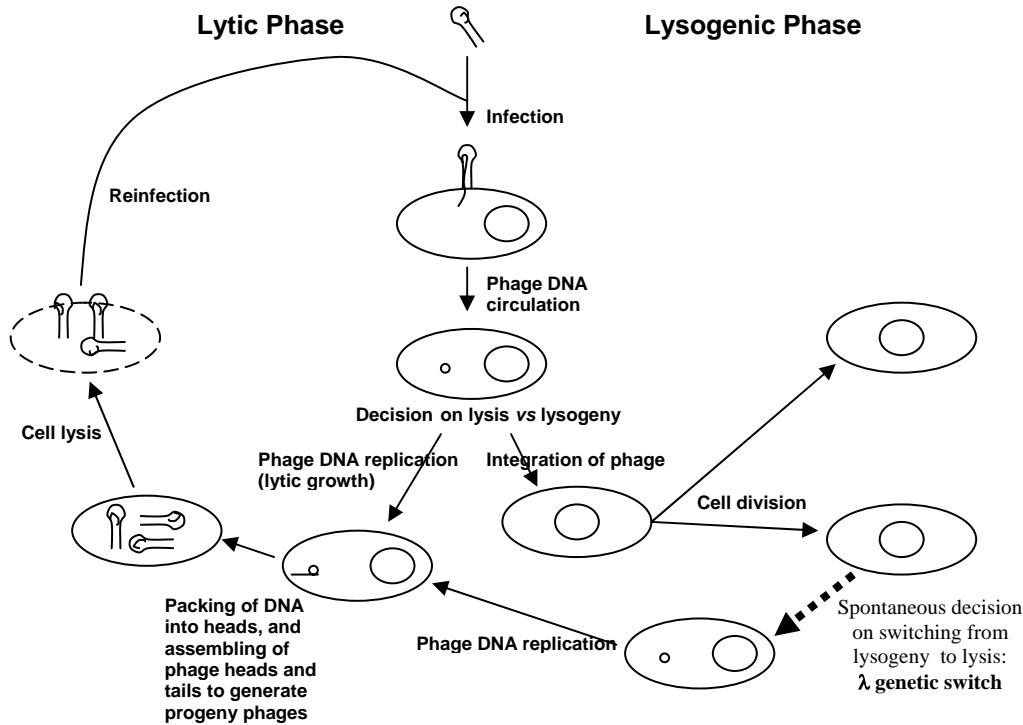

Figure 1. The schematic diagram of phage λ life cycle. The stochastic dynamics of rare spontaneous decision on switching from lysogenic phase to lytic phase is the focus of the present chapter. The induction by UV light and other SOS processes are not considered here.

## 2.2. Modeling effort

There have been continuous mathematical and numerical activities on modeling phage λ. The rationale is rather straightforward: The biological functions should emerge as the systems property from the model based on the molecular mechanism of phage regulatory elements and their independently measured parameters. The elegant physical-chemical model formulated by Shea and Ackers [48] for gene regulation of phage λ has become the base for the later studies. However, soon afterwards, Reinitz and Vaisnys [57] pointed out that the inconsistency between the theoretical results and experimental data may suggest additional cooperativity. Arkin *et al.* [58] performed stochastic simulation on phage λ development for the decision of lysogeny in the very early stage, demonstrating that this process is stochastic. Recently, Aurell and Sneppen [59] analyzed the robustness of phage λ genetic switch, using a method based Onsager-Machlup functional [58], and concluded that their theoretical analysis could not reproduce the robustness of phage λ genetic switch. Their further study confirmed their earlier results [51]. Similar puzzling was found mathematically from a different perspective [50].

The coexistence of the switch stability and switching efficiency is an apparent inconsistency for the following reasons. The lysogenic state is exceptionally stable. The fluctuations in the growth environment, the so-called extrinsic fluctuations, and the intrinsic fluctuation in the genetic switch, due to the discrete nature of chemical reactions, do not easily and accidentally flip the switch. Then when the phage is "threatened", how can the switching process become so complete with so little outside intervention? The question about internal inconsistencies in these models naturally arises: Whether the easily operable induction, or highly efficient switching, in Shea and Ackers' work [48] is a result of sacrificing the robustness of the genetic switch. Phrasing differently, if a model were so constructed that it faithfully reproduces the observed robustness of the genetic switch, whether or not it would lose the efficiency of the switch. Undoubtedly a credible model of phage λ should reproduce the properties of robustness, stability and efficiency of the genetic switch simultaneously. From such a model we should also be able to calculate the observed quantities of phage development such as the protein numbers and lysogenization frequencies. We hope to show that a foundation for such a mathematical framework against the experimental data is there, thanks to recent theoretical efforts on phage λ [9,48,57-59].

### 2.3. Modeling strategy

Our procedure is first to summarize a minimal quantitative model for the phage λ genetic switch motivated by both first principles and biological observations. We then ask the question that whether or not this minimal modeling can be successfully used to quantitatively reproduce various experimental results and is qualitatively correct in biology. If successful, the necessary modifications of molecular parameters in the modeling may be viewed as the *in vivo* and *in vitro* differences. Additional or different molecular processes inside a cell should be responsible for such differences. Some of them may be identifiable by current experimental techniques. If the answer to above question would be negative, we would conclude that the minimum quantitative modeling would not be enough. More biological causes should be looked for instead. We will show that the answer so far is positive. By combining a newly developed powerful nonlinear dynamics analysis method, which takes the stochastic force into account [61-63] and classifies the stochastic dynamical structure into four different elements, with the previously established physical-chemical model [48] a novel mathematical framework was formulated to calculate the following quantitative characteristics of epigenetic states and developmental paths [9,64]: the protein numbers in one bacterium, the protein number distributions, the lifetime of each state, and the lysogenization frequencies of mutants using the wild type as reference. We should emphasize that our review here is focused on a specific biological system, though we have made an effort to put such work in perspective.

## 3. Towards Quantitative Modeling

### 3.1. Binding configurations

The genetic switch controlling and maintaining the function of phage λ consists of two regulatory genes, *cI* and *cro*, and the regulatory regions, $O_R$ and $O_L$ on the λ DNA. Established lysogeny is maintained by the protein CI which blocks operators $O_R$ and $O_L$, preventing transcription of all lytic genes including *cro* [8,55,56]. In lysogeny the CI number functions as an indicator of the state of the bacterium: If DNA is damaged such as by the UV light the protease activity of RecA is activated, leading to degradation of CI. A small CI number allows for transcription of the lytic genes, starting with *cro*, the product of which is the protein Cro.

The decision making, or the switching, is centered around operator $O_R$, and consists of three binding sites $O_{R1}$, $O_{R2}$ and $O_{R3}$, each of which can be occupied by either a Cro dimer or a CI dimmer [55,56]. As illustrated in Fig.2, these three binding sites control the activity of two promoters $P_{RM}$ and $P_R$ for respectively *cI* and *cro* transcriptions. The transcription of *cro* starts at $P_R$, which partly overlaps $O_{R1}$ and $O_{R2}$. The transcription of *cI* starts at $P_{RM}$, which overlaps $O_{R3}$. The affinity of RNA polymerase for the two promoters, and subsequent production of the two proteins, depends on how Cro and CI bound to the three operator sites and thereby establishes lysogeny with about 500 CI molecules per bacterium. If, however, CI number becomes sufficiently small, the increased production of Cro flips the switch to lysis.

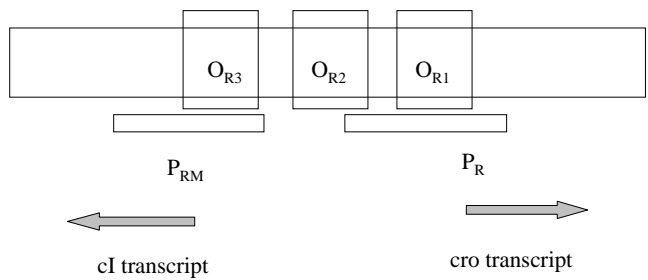

Figure 2. The $O_R$ controlling region of the phage λ genetic switch. The mathematical study [9,64] indicates that the cooperative binding of CI dimers at the $O_{R1}$ and $O_{R2}$ sites is a key to the robustness of the genetic switch. Such a cooperative binding enhances the positive CI feedback. When the CI positive feedback is turned on by the existing CI dimers, CI proteins are synthesized. The phage evolves to the lysogenic state. Otherwise Cro proteins are synthesized and the phage evolves to lytic state.

There have been numerous quantitative experimental studies on the stability in the switching of bacteriophage λ. Recently, the frequency of spontaneous induction in strains deleted for the *recA* gene has been reported independently by three groups [65-67], which was reviewed by Aurell *et al.* [67]. They all confirmed two earlier important observations that there is a switching behavior and that the switch is stable. In addition, they all obtained consistent numerical values for the switching frequency, in spite of the use of different strain backgrounds, done at different continents and at different times.

However, computational and mathematical attempts to quantitatively understand this behavior have not been successful, even permitting the possibility that the wild type may be more stable [59,67].

More recent data [68] suggest that the wild type may be two orders of magnitude more stable than previously observed value in Ref.[66]: The switching rate to lytic state may be less than $4\times10^{-9}$ per minute. In addition to the call for more experimental studies, this puts the theoretical modeling in a more challenging position. This wild type data was used as the main input to further fix the model in Ref's.[9,64]. The previous data were also discussed to illustrate a pronounced exponential sensitivity in such a modeling, summarized below.

| State | $P_{RM}$ | $O_R3$ | $O_R2$ | $O_R1$ | $P_R$ | $i(s)$ | $j(s)$ | $k(s)$ |
|---|---|---|---|---|---|---|---|---|
| 1 |  |  |  |  |  | 0 | 0 | 0 |
| 2 |  |  |  | $R_2$ |  | 1 | 0 | 0 |
| 3 |  |  | $R_2$ |  |  | 1 | 0 | 0 |
| 4 |  | $R_2$ |  |  |  | 1 | 0 | 0 |
| 5 |  |  | $R_2$ | $R_2$ |  | 2 | 0 | 0 |
| 6 |  | $R_2$ | $R_2$ |  |  | 2 | 0 | 0 |
| 7 |  | $R_2$ |  | $R_2$ |  | 2 | 0 | 0 |
| 8 |  | $R_2$ | $R_2$ | $R_2$ |  | 3 | 0 | 0 |
| 9 |  |  |  | $C_2$ |  | 0 | 1 | 0 |
| 10 |  |  | $C_2$ |  |  | 0 | 1 | 0 |
| 11 |  | $C_2$ |  |  |  | 0 | 1 | 0 |
| 12 |  |  | $C_2$ | $C_2$ |  | 0 | 2 | 0 |
| 13 |  | $C_2$ | $C_2$ |  |  | 0 | 2 | 0 |
| 14 |  | $C_2$ |  | $C_2$ |  | 0 | 2 | 0 |
| 15 |  | $C_2$ | $C_2$ | $C_2$ |  | 0 | 3 | 0 |
| 16 |  |  | $C_2$ | $R_2$ |  | 1 | 1 | 0 |
| 17 |  | $C_2$ |  | $R_2$ |  | 1 | 1 | 0 |
| 18 |  | $C_2$ | $C_2$ | $R_2$ |  | 1 | 2 | 0 |
| 19 |  | $C_2$ | $R_2$ |  |  | 1 | 1 | 0 |
| 20 |  |  | $R_2$ | $C_2$ |  | 1 | 1 | 0 |
| 21 |  | $C_2$ | $R_2$ | $C_2$ |  | 1 | 2 | 0 |
| 22 |  | $R_2$ | $C_2$ |  |  | 1 | 1 | 0 |
| 23 |  | $R_2$ |  | $C_2$ |  | 1 | 1 | 0 |
| 24 |  | $R_2$ | $C_2$ | $C_2$ |  | 1 | 2 | 0 |
| 25 |  | $C_2$ | $R_2$ | $R_2$ |  | 2 | 1 | 0 |
| 26 |  | $R_2$ | $C_2$ | $R_2$ |  | 2 | 1 | 0 |
| 27 |  | $R_2$ | $R_2$ | $C_2$ |  | 2 | 1 | 0 |
| 28 |  |  |  |  | $RNA_P$ | 1 | 0 | 1 |
| 29 |  | $R_2$ |  |  | $RNA_P$ | 1 | 0 | 1 |
| 30 |  | $C_2$ |  |  | $RNA_P$ | 0 | 1 | 1 |
| 31 | $RNA_P$ |  | $R_2$ |  |  | 1 | 0 | 1 |
| 32 | $RNA_P$ |  | $R_2$ | $R_2$ |  | 2 | 0 | 1 |
| 33 | $RNA_P$ |  | $R_2$ | $C_2$ |  | 1 | 1 | 1 |
| 34 | $RNA_P$ |  |  |  |  | 0 | 0 | 1 |
| 35 | $RNA_P$ |  |  | $R_2$ |  | 1 | 0 | 1 |
| 36 | $RNA_P$ |  |  | $C_2$ |  | 0 | 1 | 1 |
| 37 | $RNA_P$ |  | $C_2$ |  |  | 0 | 1 | 1 |
| 38 | $RNA_P$ |  | $C_2$ | $R_2$ |  | 1 | 1 | 1 |
| 39 | $RNA_P$ |  | $C_2$ | $C_2$ |  | 0 | 2 | 1 |
| 40 | $RNA_P$ |  |  |  | $RNA_P$ | 0 | 0 | 2 |

Table I. The 40 configurations corresponding to right operator (adapted from Ref's.[74-76]). $R_2$ stands for CI ($\lambda$ repressor) dimer and $C_2$ for Cro dimer.

The CI and Cro protein molecules in the cell are assumed to be in homeostatic equilibrium. There are not always the same numbers of CI and Cro dimers bound to the operators at any particular time. These numbers are fluctuating, and the equilibrium assumption should give the size of these fluctuations. The key inputs are CI and Cro dimerization constants and the Gibbs free energies for their bindings to the three operator sites $O_R1$, $O_R2$ and $O_R3$ [69-75] (See the legend of Table I and II for a more detailed description.).

Following Ackers *et al*. [76] and Aurell *et al*. [67], we encode a state s of CI and/or Cro bound to $O_R$ by three numbers (i,j,k) referring to $O_{R3}$, $O_{R2}$ and $O_{R1}$ respectively. The coding for s is 0 if the corresponding site is free, 1 if the site is occupied by a CI dimer, and 2 if the site is occupied by a Cro dimer. The probability of a state s with i(s) CI dimers and j(s) Cro dimers bound to $O_R$ is in the grand canonical approach of Shea and Ackers [48]

$$p_R(s) = Z^{-1} [CI]^{i(s)} [Cro]^{j(s)} [RNAp]^{k(s)} \exp(-\Delta G(s)/RT) \qquad Eq.(1)$$

For example, if CI occupies $O_{R1}$, Cro $O_{R2}$ and $O_{R3}$, we have i(s) = 1, j(s) = 2, k(s) = 0, and $p_R(s) = p_R(221)$. RNA polymerase (RNAp) can occupy either $O_{R1}$ and $O_{R2}$, or $O_{R2}$ and $O_{R3}$, not other configurations. There are total 40 states represented by s (Table I). The normalization constant Z is determined by summing over *s*: $Z = \Sigma_s [CI]^{i(s)} [Cro]^{j(s)} [RNAp]^{k(s)} \exp(-\Delta G(s)/RT)$. Here [ ] denotes the corresponding protein dimer concentration in the bacterium, $\Delta G(s)$ the binding energy for binding configuration s, R the gas constant and T the temperature.

### 3.2. Deterministic model

We further simplify the expression of $p_R(s)$ by noticing that protein CI and Cro control the operator [8,55,56]. If $O_{R1}$ and $O_{R2}$ are unoccupied by either CI or Cro, RNAp binds to them with a probability determined by RNAp binding energy. The case that RNAp first binds to $O_{R1}$ and $O_{R2,}$ then blocking CI and Cro binding is excluded based on the assumption that only CI and Cro controls the regulatory behavior. In addition to experimental observation, this assumption is justifiable if the time scale associated with CI and Cro binding is shorter than the RNAp binding. Except for an overall constant, which we include into the rate of transcription, the RNAp binding is no longer relevant. We therefore take it out of the expression $p_R(s)$. The total number of states is reduced to 27. This simplification was first used by Aurell and Sneppen [59]. We will drop the subscript R for binding probability $p_R$. We should point out that previous experimental and theoretical results had been concisely reviewed by Aurell *et al.* [67], whose convention we shall follow.

The dimer and monomer concentrations are determined by the formation and de-association of dimers, which gives the relation of dimer concentration to the total concentraion of proteins as

$$[CI] = [N_{CI}]/2 + \exp(\Delta G_{CI}/RT)/8$$
$$+ ([N_{CI}]\exp(\Delta G_{CI}/RT)/8 + \exp(2\Delta G_{CI}/RT)/64)^{1/2} \qquad Eq.(2)$$

Here $\Delta G_{CI} = -11.1$ kcal/mol is the dimer association free energy for CI.
Similar expression for [Cro] is

$$[Cro] = [N_{Cro}]/2 + \exp(\Delta G_{cro}/RT)/8$$
$$+ ([N_{Cro}]\exp(\Delta G_{Cro}T)/8 + \exp(2\Delta G_{Cro}T)/64)^{1/2} \qquad Eq.(3)$$

Here $\Delta G_{Cro} = -7$ kcal/mol is the dimer association free energy for Cro. Here $[N_{CI}]$ and $[N_{Cro}]$ are the monomer concentrations of CI and Cro respectively.

CI and Cro are produced from mRNA transcripts of cI and cro, which are initiated from promotor sites $P_{RM}$ and $P_R$. The rate of transcription initiation from $P_{RM}$ when stimulated by CI bound to $O_{R2}$ is denoted $T_{RM}$, and when not stimulated $T_{RM}^u$. The number of CI molecules produced per transcript is $E_{cI}$. The overall expected rate of CI production is

$$f_{CI}(N_{CI}, N_{Cro}) = T_{RM} E_{cI} [p(010) + p(011) + p(012)] +$$
$$T_{RM}^u E_{cI} [p(000) + p(001) + p(002) + p(020) + p(021) + p(022)]. \qquad Eq.(4)$$

Here $N_{CI}$ and $N_{Cro}$ are the protein numbers for CI and Cro inside the bacterium respectively. The converting factor between the protein concentration and the corresponding protein inside the bacterium is listed in Table II. Similarly, the overall expected rate of Cro production is

$$f_{Cro}(N_{CI}, N_{Cro}) = T_R E_{cro} [p(000) + p(100) + p(200)]. \qquad Eq.(5)$$

We use $T_{RM}$, $E_{cI}$, $E_{cro}$, and $T_{RM}^u$ from Aurell and Sneppen [59], which were deduced from the resulting protein numbers in lysogenic and lytic states.

The free energies $\Delta G(s)$ are determined from *in vitro* studies, that is, they are obtained outside of the living bacterium. The *in vivo* conditions for inside a living bacterium could be different. The measured protein-DNA affinities could depend sensitively on the ions present in the buffer solutions as well as other factors. This observation will be important in our comparison between theoretical results and experimental data. On the other hand, the *in vivo* effects of such changes should be compensated for, as e.g. changed KCl concentrations are by putrescine [77] and other ions and crowding effects [78]. We note that Record *et al.* [78] already observed that there may exist a significant difference between *in vivo* and *in vitro* molecular parameters. The data quoted in Darling *et al.* was obtained at KCl concentration of 200mM, which resembles *in vivo* conditions. Therefore, though we expect a difference between the *in vivo* and *in vitro* data, the difference may not be large, typically within 20-30% of the *in vitro* values.

The mathematical model which describes the genetic regulation in Fig.2 is a set of coupled equations for the time rate of change of numbers of CI and Cro in a cell [57]:

$$dN_{CI}(t)/dt = F_{CI}(N_{CI}(t), N_{Cro}(t))$$
$$dN_{Cro}(t)/dt = F_{Cro}(N_{CI}(t), N_{Cro}(t)) \qquad \text{Eq.(6)}$$

where the net production rates are

$$F_{CI} = f_{CI}(N_{CI}, N_{Cro}) - N_{CI}/\tau_{CI}$$
$$F_{Cro} = f_{Cro}(N_{CI}, N_{Cro}) - N_{Cro}/\tau_{Cro} \qquad \text{Eq. (7)}$$

Eq.(6) and (7) represent the minimum deterministic model. Here dN/dt is the rate N changes. The production terms $f_{CI}$ and $f_{Cro}$ are functions of CI and Cro numbers in the bacterium. With no Cro in the system, the curve of $f_{CI}$ vs. CI number has been experimentally measured [79]. As reviewed in Aurell *et al.* [67] these measurements are consistent with the best available data on protein-DNA affinities [69,71,80] dimerization constants[81], initiation rates of transcriptions of the genes, and the efficiency of translation of the mRNA transcripts into protein molecules. The decay constant $\tau_{CI}$ is an effective life time, proportional to the bacterial life-time, since CI molecules are not actively degraded in lysogeny, while $\tau_{Cro}$ is about 30% smaller [82]. We comment that there is considerably more experimental uncertainty in the binding of Cro, both to other Cro and to DNA, than the binding of CI, see e.g. Darling *et al.* [74,75]. As a minimal mathematical model of the switch, we take $\tau_{CI}$ and $\tau_{Cro}$ from data, and deduce $f_{CI}$ and $f_{Cro}$ at non-zero number of both CI and Cro with a standard set of assumed values of all binding constants, which are summarized by Aurell *et al.* [67] and are adopted here (Table II with differences in cell volume and converting factor, as well as the *in vivo* and *in vitro* differences)

**3.3. Positive and negative feedbacks; *in vivo* vs *in vitro***

Both positive and negative feedbacks are employed in this genetic switch. For CI (Cro), it has a positive feedback effect on itself and a negative feedback effect on the production of Cro (CI) [8]. Evidently, these feedbacks are systems effects: breaking them down into disintegrated parts the feedback effects would disappear. They emerge only when a proper integration is done. Such a systems effect is well known in engineering [83].

In establishing the minimum deterministic model, another major implicit assumption is on the time scales. We have assumed that the dimerization process is a fast process on the scale of Cro and CI production, hence can be treated as algebraic constraints. The dimerization has been subjected to continuous experimental [84] and theoretical [85] studies. It was concluded by *in vitro* experiment [84] that the Cro dimerization is slower than that of CI. Nevertheless, the Cro dimerization time is on the order of fraction of a minute [84], which is much smaller than the typical time of order of 20 minutes used in our modeling (*c.f.* Table II). We may be able to apply the useful quasisteady state approximation [86]. Thus, the algebraic constraints appears to a reasonable assumption for such a minimum modeling used in [9,59,64,67]. Other cellular processes have also been implicitly assumed to be fast. All their residual effects will be treated as extrinsic stochastic effect contribution, and will be incorporated into the minimum modeling in the name of intrinsic *vs* extrinsic noises, to be discussed below.

We specify further the meaning of minimum deterministic modeling. First of all, such a model should be viewed as what the system might be, not as what it must be. Many features are not explicitly contained in it, such as the nonspecific binding [87] and the looping [88,89]. The non-specific binding was already demonstrated to be not crucial, but the looping may well be, to which we will come back later. Nevertheless we point out that by assuming the minimum deterministic modeling, we tentatively and tactically assume it has captured all the essential features of the λ switch by aggregating molecular processes around it. In do so it suggests another understanding of the difference between *in vivo* and *in vitro*: All the parameters we adopted from *in vitro* measurement would indeed take a different value *in vivo*, because there exist numerous other biological processes inside the cell which contribute to this difference. If the minimum model is essentially correct, it should be able to account for experimental data in a quantitative manner, along with predictions to be further tested. We will show below that we have indeed achieved this goal after several decades of theoretical efforts.

## 4. Stochastic Dynamical Modeling

### 4.1. Minimum quantitative model

Stochasticity is ubiquitous in biology. For the present modeling, it is particularly easy to motivate it. If the numbers of CI and Cro were macroscopically large, then Eq.(1) would be an entirely accurate description of the dynamics, because the fluctuation in number is an order of $N^{1/2}$ and the correction is an order of $1/N^{1/2}$, which would be negligibly small when N would be very large. The numbers are however only in the range of hundreds. Hence the fluctuation is not negligible. The actual protein production process is influenced by many chance events, such as the time it takes for a CI or a Cro in solution to find a free operator site, or the time it takes a RNA polymerase molecule to find and attach itself to an available promoter, suggesting more stochastic sources. As a minimal model of the network with finite-N noise, we therefore consider the following system of two coupled stochastic differential equations, with two independent standard Gaussian and white noise sources:

$$dN_{CI}/dt = F_{CI} + \zeta_{CI}(t)$$
$$dN_{Cro}/dt = F_{Cro} + \zeta_{Cro}(t) \qquad \text{Eq.(8)}$$

We further assume that the means of the noise terms are zero, *i.e.* $<\zeta_{CI}(t)> = <\zeta_{Cro}(t)> = 0$, with the variance

$$< \zeta_{CI}(t) \zeta_{CI}(t') > = 2 D_{CI} \delta(t-t')$$
$$< \zeta_{Cro}(t)\zeta_{Cro}(t') > = 2 D_{Cro} \delta(t-t')$$

$$<\zeta_{CI}(t)\, \zeta_{Cro}(t')> \ = 0 \qquad \text{Eq.(9)}$$

Eq.(8) and (9) consist of the present minimum quantitative model. Here the symbol $< \ldots >$ denotes the average over noise. Eq.(9) defines a 2×2 diffusion matrix $D$. The noise strength may contain contributions from the production and decay rates, assuming each is dominated by one single independent reaction, as used by Aurell and Sneppen [59]. Such a noise may be called the `intrinsic' noise. Other noise sources, 'extrinsic' noises, also exist [90-93]. We treat the noise to incorporate both intrinsic and extrinsic sources: All are assumed to be Gaussian and white. The consistent of this assumption should be tested experimentally, as will be the case below. Certain probability events, however, may not behave as Gaussian and white in the present context of modeling, which can be determined by separate biological experiments, such as the $p_{RM}240$ mutation [68] to be discussed below.

It has been demonstrated [61-63] that there exists a unique decomposition such that the stochastic differential equation, Eq.(8), can be transformed into the following form, the four dynamical element structure:

$$[\Lambda(\mathbf{N}) + \Omega(\mathbf{N})]\, d\mathbf{N}/dt = -\nabla U(\mathbf{N}) + \boldsymbol{\xi}(t) \qquad \text{Eq.(10)}$$

with the semi-positive definite symmetric 2×2 matrix $\Lambda$ defining the dissipation (degradation), the anti-symmetric 2×2 matrix $\Omega$ defining the transverse force, the single valued function U defining the potential landscape, and the noise vector $\boldsymbol{\xi}(t)$, and the two dimensional vectors:

$$\begin{aligned}
\mathbf{N}^\tau &= (N_{CI}, N_{Cro}); \\
\nabla &= (\partial/\partial N_{CI},\ \partial/\partial N_{Cro}); \\
\boldsymbol{\xi}^\tau &= (\xi_{CI}, \xi_{Cro}),
\end{aligned} \qquad \text{Eq.(11)}$$

here $\tau$ means the transpose of the vector. The connection between the noise $\boldsymbol{\xi}$ and the matrix $\Lambda$ is similar to that of $\zeta$ and $D$ of Eq.(9):

$$\begin{aligned}
<\boldsymbol{\xi}(t)> &= 0 \\
<\xi_{CI}(t)\, \xi_{CI}(t')> &= 2\, \Lambda_{CI}\, \delta(t-t') \\
<\xi_{Cro}(t)\, \xi_{Cro}(t')> &= 2\, \Lambda_{Cro}\, \delta(t-t') \\
<\xi_{CI}(t)\, \xi_{Cro}(t')> &= 0
\end{aligned} \qquad \text{Eq.(12)}$$

The decomposition from Eq.(8) and Eq.(9) to Eq.(10) and Eq.(12) is determined by the following set of equations:

$$\nabla \times [(\Lambda + \Omega)\, \mathbf{F}] = 0 \qquad \text{Eq.(13)}$$

$$(\Lambda + \Omega) D (\Lambda - \Omega) = \Lambda. \qquad \text{Eq.(14)}$$

One may solve for $\Lambda$, $\Omega$ in terms of $\mathbf{F} = (F_{CI}, F_{Cro})^\tau$ and $D$ from Eq.(13) and Eq.(14). Indeed, this can be formally done. Once $\Lambda$, $\Omega$ are known, the requirement that Eq.(10)

can be reduced to Eq.(8) gives $(\Lambda + \Omega) \mathbf{F} = -\nabla U(\mathbf{N})$, which is used to obtain U. In general this decomposition is an involved mathematical and numerical endeavor. Further simplification follows from the simplification of friction matrix. Typically the diffusion matrix *D* is unknown biologically: There are no enough measurements to fix the noise explicitly. Therefore we may treat the semi-positive definite symmetric matrix $\Lambda$ as parameters to be determined experimentally. In our calculation, we then assume that *D* is a diagonal matrix. Following from Eq.(14), $\Lambda$ is a diagonal matrix for two dimensional case. The experimentally measured fraction of *recA*$^{-1}$ lysogens that have switched to lytic state is used to determine the elements of $\Lambda$.

Here we would like to give an intuitive interpretation of the mathematical procedure. Eq.(8) corresponds to the dynamics of a fictitious massless particle moving in two dimensional space formed by the two protein numbers $N_{CI}$ and $N_{Cro}$, with both deterministic and random forces. It is easy to check that in general $\nabla \times \mathbf{F}(\mathbf{r}) \neq 0$ and $\nabla \cdot \mathbf{F}(\mathbf{r}) \neq 0$. Therefore $\mathbf{F}(\mathbf{r})$ cannot be simply represented by the gradient of a scalar potential due to both the force transverse to the direction of motion and force of friction. The simplest case in two dimensional motion when both transverse force and friction exist is an electrically charged particle moving in the presence of both magnetic and electric fields, which is precisely in the form of Eq.(10).

Proceeding from Eq.(10), we note that we may interpret the semi-positive definite symmetric $\Lambda$ matrix as the friction matrix, and the antisymmetric matrix $\Omega$ as the result of a `magnetic' field. The friction matrix represents the dissipation in physics. It is analogous to the degradation in biology. The scalar function U takes the role of a potential function which would determine the final steady distribution of the phage. The global equilibrium will be reached when the final distribution function is given by

$$\rho(N_{CI}, N_{Cro}) = \exp(-U(\mathbf{N}))/\int dN_{CI} \int dN_{Cro} \exp(-U) \qquad \text{Eq.(15)}.$$

The potential U, the landscape of the system, is depicted in Fig. 2 (*c.f.* Fig. 4).

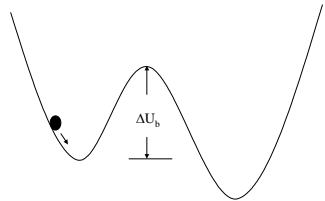

Figure 3. Illustration of the dynamical structure of a genetic switch. The dynamic state of the network is represented by a particle whose position is given by instantaneous protein numbers. The potential function maps a landscape in the protein number space. For a genetic switch, there are two potential minima corresponding to two epigenetic states. The area around each of the minima forms the attractive basin. The state of the network always tends to relax to one of the minima. The fluctuation may bring the network from one minimum to another with a rate given by Kramers rate formulae [94,95].

The phage sees two minima and one saddle point in the potential landscape. Those two minima correspond to the lytic and lysogenic states. Once the phage is at one of the minimum, the probability rate for it to move into another minimum is given by the Kramers rate formulae in the form [94,95]:

$$P = \omega_0 \exp(-\Delta U_b) \qquad \text{Eq.(16)}$$

with the potential barrier height $\Delta U_b = (U_{saddle} - U_{initial\ minimum})$, the difference in potential between the saddle point and the initial minimum, and the time scale, the attempt frequency $\omega_0$, determined by the friction, the curvatures of potential and values of transverse force around the saddle and the local minimum. We remark here that the attempt frequency is in general a complicated function of dynamical quantities in Eq,(10). It's form will be determined empirically in the present paper. We refer readers to Ref.[95] for the general mathematical discussions.

**4.2. Stochastic dynamical structure analysis**

Eq.(10) gives the dynamical structure of the gene regulatory network in terms of its four dynamical components: the friction, the potential gradient of the driving force, the transverse force, and the stochastic force. Such a dynamical structural classification serves two main purposes. It provides a concise description for the main features of the genetic switch by itself and it provides a quantitative measure to compare different gene regulatory networks, for instance between the wild phage and its mutant. Such an analysis of experimental data based on Eq.(10) will be tentatively named the dynamical structure analysis.

The potential may be interpreted as the landscape map of the phage development. Each of the epigenetic state is represented by a potential minimum and its surrounding area forms an attractive basin. The dissipation represented by the friction gives rise to the adaptivity of the phage in the landscape defined by the potential: The phage always has the tendency to approach the bottom of the nearby attractive basin. The potential change near the minimum, together with the friction, gives the time scale of relaxation: The time it takes to reach equilibrium after the epigenetic state is perturbed. Once we know the friction and the potential around the minimum, we have a good grasp of the relaxation time, $\tau = \eta/U''$, here $\eta$ is the strength of friction, $U''$ the second derivative of potential, both in one dimensional approximation along a relevant axis. The relaxation time is independent of the amplitude of the perturbation near the potential minimum, when $U''$ is a constant.

Two remarks are in order here. 1. The meaning of friction matrix is the same as in mechanics: If there is no external driving force, the system tends to stop at its nearby minimal position. The closest corresponding concept in biology is `degradation': There is always a natural protein state under given condition. 2. It turns out that the transverse force is not a dominant factor in the present switch-like behavior. However, its existence is the necessary condition for oscillatory biological behaviors, which will not be discussed further in the present paper.

Another time scale provided by the potential is the lifetime of the epigenetic state, which is given by the Kramers rate formulae, Eq. (16), through the potential barrier height. Such a scale measures the stability of epigenetic state in the presence of fluctuating environment. In the case of phage λ, the lifetime for lysogenic state is very long, unless the phage is mutated at its operator sites. When the phage is provoked, the height of the potential barrier separating lysogenic and lytic states is reduced. The lifetime of lysogenic state is drastically reduced due to its exponential dependence on the barrier height and switching takes place. Looking at it from a different angle, the stochastic force gives the phage ability to search around the potential landscape by passing through saddle points and drives the switching event. The Kramers rate formulae is a quantitative measure of this optimization ability.

## 5. Quantitative Comparison between Theory and Experiment

### 5. 1. Determining *in vivo* parameters.

First we need to decide the free energies to be used in the theoretical model. Without exception, all the binding energies measured so far for phage λ are determined from *in vitro* studies. The difference between the *in vivo* condition and the *in vitro* condition could include the ion concentration in the buffer solutions and the spatial configuration of the genomic DNA, for instant looping [96-98]. The relative large change of the cooperative energy from *in vitro* to *in vivo* in Table I may be partly due to the looping effect, though there is no direct consideration of looping in present model. We note that in the *in vivo* conditions all the operators are in the same kind of environment, including the ion condition and the DNA configuration. The reason for the latter is that the operators are closely located to each other in the genome. If there is a bending of the genomic DNA which increase or decrease DNA-protein bindings, these closely located and short operator sites are mostly likely experience the same amount of change. Therefore, we assume that in addition to the *in vitro* DNA-protein binding energy, overall binding energy differences are added to all the CI and Cro protein respectively:

*in vivo* binding energy for CI (Cro) = *in vivo* binding energy for CI + $\Delta G_{CI}$ ($\Delta G_{Cro}$).

To determine $\Delta G_{CI}$($\Delta G_{Cro}$), we need more experimental input than the *in vitro* measurement. To avoid unnecessary uncertainty in the model, we try to include a minimal number of parameters. The cooperative binding between two CI dimers is included. The cooperative bindings between two Cro dimers, between CI and Cro dimers, the unspecific CI and Cro bindings are not included. Our later calculation verifies that CI cooperative binding is essential to the genetic switch properties while the bindings we ignore do not have significant influence on the calculated results. There are three parameters we need to adjust, the difference between *in vivo* and *in vivo* binding energy for CI ($\Delta G_{CI}$), for Cro ($\Delta G_{Cro}$), and for the cooperativity of CI dimers ($\Delta G$(cooperative)). We first use the CI numbers of both wild type and mutant $\lambda O_R 121$ to determine $\Delta G_{CI}$, then we determine $\Delta G$(cooperative) and $\Delta G_{Cro}$ by requiring both the lytic and lysogenic states of wild type are equally stable, calculated from Kramers rate formulae. The adjusted *in vivo* binding energies and other parameters we use for the modeling are given in Table I. Using these adjusted parameters the robustness of the phage's genetic switch is reproduced (shown in Fig. 4).

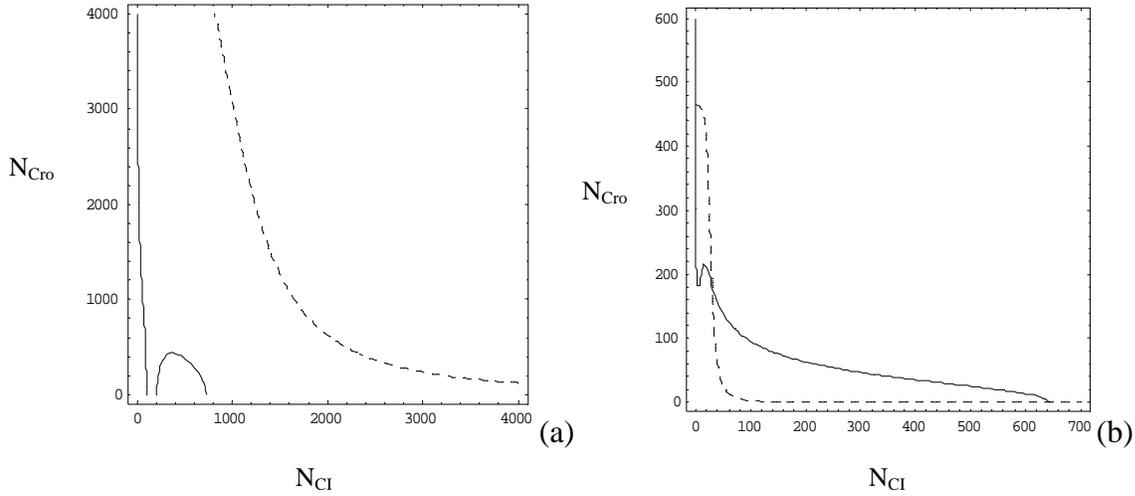

Figure 4. Lines of $d<N_{CI}>/dt = 0$ (solid) and $d<N_{Cro}>/dt = 0$ (dashed), here $<>$ is the average to stochastic force, for: (a) the wild type phage, $\lambda O_R 321$, with parameters taken directly from *in vitro* measurement; (b) the wild type phage with parameters adjusted allowing *in vivo* and *in vitro* differences. For mutants $\lambda O_R 121$ $\lambda O_R 323$ see Ref.[9]. For (b) these two lines have three intersections. These three fixed point in Eq.(2) coincide with the potential extrema, minima and saddle point, in Eq.(10).

The mutant $\lambda O_R 3'23'$ studied by Little *et al*. [66] was characterized by Hochschild *et al*. [99] for binding to $O_R 3'$. To produce the desired protein level, we obtained that the binding energy between $O_R 3'$ and Cro protein is 1.8 kcal/mol smaller than that of the $O_R 3$ and Cro protein, consistent with the result of Hochschild *et al*. The CI binding energy from $O_R 3$ to $O_R 3'$ is slightly increased, 1 kcal/mol, also consistent with the measurement.

| | |
|---|---|
| RT | 0.617 kcal/mol |
| Effective bacterial volume | $0.7 \times 10^{-15}$ $l$ |
| $E_{cro}$ | 20 |
| $E_{cI}$ | 1 |
| $T_{RM}$ | 0.115 /s |
| $T_{RM}^u$ | 0.0105 /s |
| $T_R$ | 0.30 /s |
| $\tau_{CI}$ | $2.9 \times 10^3$ s |
| $\tau_{Cro}$ | $5.2 \times 10^3$ s |
| Converting factor between protein number and concentration | $1.5 \times 10^{-11}$ |
| *in vitro* free energy differences for wild type $\lambda$ $\Delta G (001)$ $\Delta G (010)$ $\Delta G (100)$ | -12.5 kcal/mol -10.5 kcal/mol -9.5 kcal/mol -25.7 kcal/mol |
| ΔG (011) | -22.0 kcal/mol |
| ΔG (110) | -35.4 kcal/mol |
| ΔG (111) | -14.4 kcal/mol |
| ΔG (002) | -13.1 kcal/mol |
| ΔG (020) | -15.5 kcal/mol |
| ΔG (200) | -2.7 kcal/mol |
| ΔG(cooperative) dimerization energy | |
| ΔG$_{CI2}$ | -11.1 kcal/mol |
| ΔG$_{Cro2}$ | -7.0 kcal/mol |
| *in vitro* free energy differences for O$_R$3' binding | |
| ΔG (100) | -10.5 kcal/mol |
| ΔG (200) | -13.7 kcal/mol |
| *in vivo* free energy differences - *in vitro* free energy differences | |
| ΔG$_{CI}$ | –2.5 kcal/mol |
| ΔG$_{Cro}$ | -4.0 kcal/mol |
| **ΔG(cooperative)** | **-3.7 kcal/mol** |

Table II. Parameters used in the modeling. CI dimer affinities to O$_{R1}$, O$_{R2}$ and O$_{R3}$ are after Darling *et al.* [74,75]. Cro dimer affinities to O$_{R1}$, O$_{R2}$ are after Takeda *et al.* [70,80], Jana *et al.* [73], Kim *et al.* [71], as stated in Aurell *et al.* [67]. The CI dimerization energy is from Koblan and Ackers [81], Cro dimerization energy from Jana *et al*. [72,73]. The bacterial volume is taken from Bremmer and Dennis [100]. E$_{CI}$ and E$_{Cro}$ are taken from Shean and Gottesman [101], Ringquist *et al.* [102], and Kennell and Riezman [103.] The *in vitro* parameters have been summarized by Aurell *et al.* [67], which we largely follow. However, we here point out two differences: 1) Our effective bacterial volume, estimating from the typical size of the bacterium, assuming a tube of about 0.7micrometer in diameter and 2 micrometers in length, is about factor 3 smaller; and 2) The normalization factor for the concentrations, calculated against the numbers of water molecules, is about factor of 60 smaller. The difference between *in vivo* and *in vitro* values is consistent with qualitative experimental observation [8]. The wild type data of Little [68] is used to determine the *in vivo* and *in vitro* difference.

We assume that friction matrix $\Lambda$ is a diagonal constant matrix. Similar to Aurell and Sneppen [59], we assume the stochastic fluctuations in Eq.(2) scale with the square root of protein number divided by relaxation time: $D_{CI} = \text{Const} \times \tau_{CI}/N_{CI,lysogen}$, and $D_{Cro} = \text{Const} \times \tau_{Cro}/N_{Cro,lysis}$, where N$_{CI,lysogen}$ is the CI number at lysogenic state and N$_{Cro,lysis}$ is the Cro number at the lytic state. The Const is to be determined by experiments. In Eq.(7), we note that if the antisymmetric matrix $\Omega$ is small, that is, |det ($\Omega$)| << det($\Lambda$) then $\Lambda$ is the inverse of *D*. We calculate $\Omega$ assuming $\Lambda = D^{-1}$ and find that indeed in the regions of concern, i.e. the potential valley connecting two potential minima through the saddle points, $\Omega$ is negligible. The final parameters we have used are

$$\Lambda_{11} = 0.056 \times \tau_{CI}/N_{CI,lysogen}$$
$$\Lambda_{22} = 0.040 \times \tau_{Cro}/N_{Cro,lysis} \ .$$  Eq.(17)

**5.2. Stochastic dynamical structure analysis of λ switch**

The original problem, described by Eq.(8), may be interpreted as a set of two dimensional differential equation describing a particle motion, if we view the protein number $N_{CI}$ and $N_{Cro}$ as the coordinates and the particle position to be ($N_{CI}$(t) $N_{Cro}$(t)) at time t. There is a deterministic force $\mathbf{F}^\tau = (F_{CI}, F_{Cro})$ and a stochastic force acting on such a particle. The deterministic force has the characteristics of a friction, a potential force, and a transverse force at the same time. The decomposition we have discussed earlier, Eq.(10) allows to separate these components. We discuss them respectively here.

The wild type phage λ and some of its mutants sees two minima and one saddle point in the potential energy landscape (Fig. 5). Those two minima correspond to the lytic and lysogenic states (*cf.* Fig. 3). The positions of the potential minima give the average protein number for lytic and lysogenic states. There is a relatively narrow valley connecting these two minima. The highest point along this valley is the saddle point. Since the areas with large potential are not easily accessible and the low-lying potential region forms a valley, we may visualize the potential along the valley and illustrate it in a one dimensional graph as shown in Fig.3 and Fig.6.

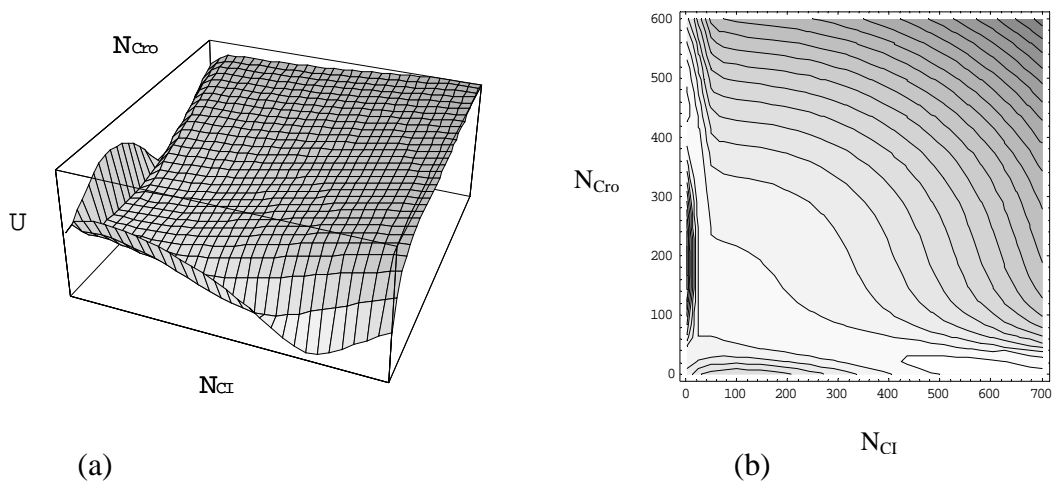

(a)      (b)

Figure 5. The potential U of wild type phage plotted on logarithmic scale (a) and as a contour map (b). There are two potential minima corresponding to the lysogenic and lytic states. Connecting these two states is a narrow potential valley. The highest point along this valley is the saddle point. The most probable state of the phage is at either of the potential minima. The fluctuation may bring the phage from the original potential minimum, moving along the valley and across the saddle point to reach another potential minimum. The rate for such a switching event is given by the Kramers rate formulae.

The antisymetric matrix $\Omega$ may be represented by a single scalar B along the *z* direction $\Omega \mathbf{F} = B\ z \times \mathbf{F}$, assuming $x = N_{CI}$, $y = N_{Cro}$. The transverse field B for the wild type is obtained by numerically solving Eq.(13). This field is small except at the region along two axes. Along the two axes, the transverse B field has no effect, since the motion is guided by the steep potential to a valley. Once the phage evolves away from origin when both CI and Cro number are small, in the later development the transverse force may be taken out of Eq.(10) without changing the dynamics of the phage. In both the calculation of the relaxation time and the lifetime of lysogenic state, we may ignore the transverse force for the above reason.

The protein number distributions of Cro and CI has also been calculated in Ref.[9]. We refer readers there for details as well as for the analysis of other quantities, such as the robustness and stability.

**5.3. Switch efficiency**

Efficiency is an important feature which so far has received relative less attention in literature. We present the discussion in some detail here.

The analysis of robustness of phage λ genetic switch demonstrates that its epigenetic states are stable against the variations in parameters and robust against major changes in terms of mutations. Then how does the switching take place? From theoretical point of view, there are two channels that the phage can be induced from lysogenic growth to lytic growth. In reality phage seems to use both of these strategies. For clarity, we begin by discussing these two channels separately.

The first channel of induction is to increase the noise level of CI protein number while keeping all the other conditions intact. Mathematically it means to increase $\zeta_{CI}$ in Eq.(8) and $D_{CI}$ in Eq.(9), while keeping all the other terms in Eq.(8) and Eq.(9) unchanged. The friction matrix $\Lambda$ is changed through the decomposition procedure. As a result, the potential energy U is also changed. Therefore for different noise level, the phage moves in different potential landscape. Such a change of noise level has a drastic effect. It changes the minima of the potential well of lysogen by making it shallower. As a good approximation, the barrier height of the lysogen potential well scales inversely with the noise strength. Doubling the noise level reduces the potential barrier by half. As a result, the increased noise level drastically decreases the lifetime of lysogenic state, as shown in Fig. 10. The lifetime of the lytic state, on the other hand, remains unchanged. The combination of these two changes in the potential landscape brings the phage to lytic growth efficiently.

The second channel is through the deterministic terms in Eq.(8). For the deterministic terms, for example, introducing CI monomer cleavage is equivalent to substitute $N_{CI}$ in Eq.(8) with $\alpha N_{CI}$, here α is a factor represent cleavage strength (Fig. 11). If α is smaller than 0.02, we find that lysogenic state is no longer stable, i.e. no longer a potential minima. The interpretation of such a small α is that almost each of the CI monomer is cleaved. If α is small, say 0.1, meaning 90% CI monomers are cleaved, the lysogenic state is still stable with a lifetime almost unchanged. Apparently a uniform CI cleavage alone without introducing extra noise to CI levels is not an efficient way for induction.

Phage may have used both of these two channels. The second channel is obviously used, since RecA cleaves CI monomers. The strong indication that the first channel is also used comes from the observations that without external stimulus, the *recA*[+] phage shows a much shorter lifetime for the lysogenic state compared to *recA*[-] phage. Such a significant reduction of lysogen lifetime without activating RecA proteins on an observable scale can be explained by doubling of the CI noise level. Fig.6 gives schematic explanations of the switching process.

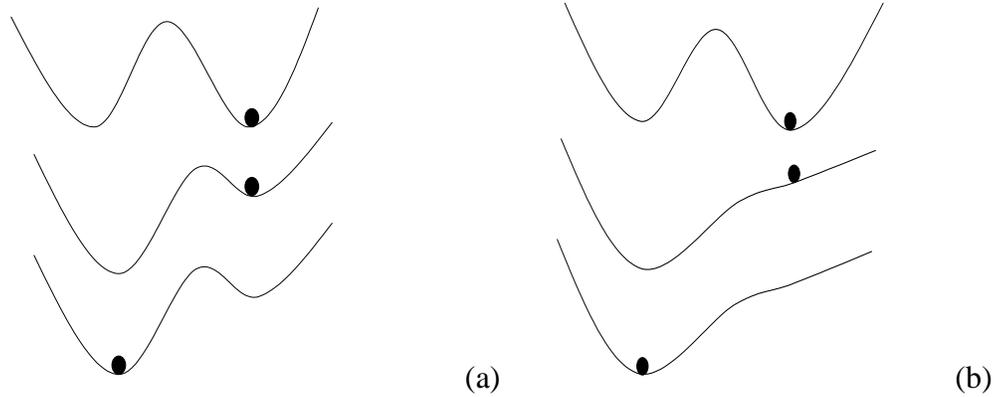

Figure 6. (a) Illustration of the switching mechanism from the current work. Before switching the phage grows in lysogenic state. The potential barrier separating the lysogenic state and the lytic state is high. When *recA* is activated, this barrier is lowered. The lifetime of lysogenic state reduces drastically and the phage switches to lytic state. (b) Switching mechanism of Shea and Ackers [48]. In their work fluctuation was not included. Switching was only possible when the lysogenic state is no longer a potential minimum. When stochastic effect is included, the switching happens when the lysogenic potential minimum become too shallow to confine fluctuation.

In the early work by Shea and Ackers [48], stochastic effect was not included. In their model even a shallow lysogenic potential minimum would confine the phage to continue growing in lysogenic state. Switching happens only when the lysogen potential minimum disappears completely. For the parameters they used, they find that when 20% of CI monomers were cleaved, such a switching would happen. As pointed out by Aurell *et al*. [67], in these early works the genetic switch modeling results do not show the observed robustness. After we require that the genetic switch should demonstrate the observed robustness, the disappearance of the lysogenic potential minimum is pushed down to 2%. However, the actually switching happens before the disappearance of lysogen potential minimum when the lysogen potential minimum is too shallow to confine the fluctuations. If 10% of CI monomer is cleaved, at least we expect a 10-fold increase in $D_{CI}$ due to the reduced CI monomer numbers. The potential barrier for lysogenic state reduces to less than 1, therefore becomes too shallow to allow continual lysogenic growth.

**5.4. Quantitative comparison with experiment of Little *et al*.**

We summarize the calculation results related to measurement of Little *et al*. [66] in Table III since their data are most update and systematic. In their experiment, they measured the free phage per lysogenic cell for both *recA$^+$* and *recA$^-$* phage, but did not convert the *recA$^+$* into fraction of lysogens that switched to lytic state. If we assume the burst size for both the *recA$^+$* and the *recA$^-$* phage are similar, our calculation for the RecA$^+$ protein agrees with their measurements quantitatively.

| Phage genotype | Relative CI level in lysogen | Relative Cro level in lysis | Switching frequency to lytic state ($recA^-$) per minute | Switching frequency to lytic state ($recA^+$) per minute |
|---|---|---|---|---|
| | Theoretical (experimental) | Theoretical | Theoretical (experimental*) | Theoretical |
| $\lambda^+$ | 100% (100%) | 100% | $1\times10^{-9}$ ($2\times10^{-9}$) | $1\times10^{-5}$ |
| $\lambda O_R 121$ | 20% (25-30%) | 100% | $3\times10^{-6}$ ($3\times10^{-6}$) | $3\times10^{-5}$ |
| $\lambda O_R 323$ | 70% (60-75%) | 70% | $7\times10^{-5}$ ($2\times10^{-5}$) | $1\times10^{-4}$ |
| $\lambda O_R 3'23'$ | 50% (50-60%) | 130% | $1\times10^{-7}$ ($5\times10^{-7}$) | $2\times10^{-5}$ |

Table III. Comparison between the calculation and the experiment data (in parentheses) by Little *et al*. [66] Here *) indicates that the estimated wild type data from Little (private communication) [68] is used. The wild type biological data were used to find out the difference between *in vivo* and *in vitro* molecular parameters, as listed in Table I. The relative CI level and switch rate of $\lambda O_R 121$ were used to fine tune parameters. Rest theoretical entries are then calculated directly from our model.

In Table III the bi-stability of the gene switch in phage λ is assumed, and the protein levels in the lytic state are calculated. This is of course not the case for the wild type, hence posts a question to test the calculated Cro level experimentally. One way to realize the bi-stability may be by suppressing the lyses, achieving the so-called anti-immune phenotype [104,105].

As discussed in section IV of formulation of the present stochastic model, we have made the simplified assumption of treating all chance or probability events as Gaussian white noise. This assumption affects two testable biological quantities: the lifetime of lysogenic state, Eq.(16), and the shape of CI number distribution in lysogenic state, Eq.(15) and Fig. 7. Simultaneous measuring both of them can be used as a consistent check to the Gaussian white noise assumption. For example, we have treated the effect of $recA^+$ to switching dynamics as that of a Gaussian white noise to simplify our calculation, in the same reasoning of minimal modeling type approach in the present paper. In fact, we have assumed that with $recA^+$ the total noise strength doubles. Using this assumption we calculated the lytic switching rates, represented by the last column of Table III. The CI distribution with $recA^+$ should be twice as broad as in the case with $recA^-$. Both results are subjected to the further experimental test.

There may be some chance events that cannot be treated as Gaussian white noise in the present formulation. One example has been already suggested in biological experiments [68], the $p_{RM}240$ mutation which greatly weakens the promoter therefore the ability to produce CI. This mutation makes the lysogens barely stable, and is estimated to responsible for at least 99% of observed lytic switching in the wild type. We have used this input for both Table II and III. We have re-calculated the switching rates to lytic state of all strands, assuming the same minimal model, with the same forms of functions for the switching rate, but with the previous experimental data [66]. The switching rates obtained in this way are: wild type($\lambda^+$), $2\times10^{-7}$; $\lambda O_R 121$, $\lambda O_R 2\times10^{-6}$; $\lambda O_R 323$, $7\times10^{-5}$; $\lambda O_R 3'23'$, $5\times10^{-7}$. Indeed, the stability of the wild type decreases by more than 2 orders of magnitude. The overall noise strength is increased by 60% for the wild type, resulting a broader CI distribution in lysogenic state. There is no appreciable change in other quantities, such as the protein level. The only noticeable overall change in molecular parameters is the *in vivo* cooperative energy, from –6.4 kcal/mol to –6.7 kcal/mol. A good overall quantitative agreement exists between modeling and experiment.

It is a fact that any mathematical modeling in natural science should have empirical input to completely fix its mathematical structure. For the modeling of phage λ there is an already large body of molecular data which enables us to nearly pin down our model. The additional freedom in our parameters is fixed by data from wild type, such as the switching frequency. Above less-than-expected sensitivity of our mathematical structure to this frequency that a few percentage of change in molecular parameters can result in two orders of magnitudes change in frequency is a remarkable demonstration of the internal consistence of our modeling. It demonstrates that the switching is exponentially sensitive to some molecular parameters. In addition to more theoretical effort to go beyond our present minimal modeling, it is clear that more experiments are needed in this direction to test the present model: The precise *in vivo* molecular parameters and the distributions and time-correlation of protein numbers in our model should be viewed as predictions.

**5.5. Experimental determination of dynamical elements**

We have introduced four dynamical quantities for a gene regulatory network: friction, potential, the transverse force and the stochastic force. The friction and the strength of the stochastic force are related. For the genetic switch, the transverse force is irrelevant to the dynamic properties. Therefore the two crucial quantities for a genetic switch are the friction and the potential. Those quantities can be calculated from the more microscopic modeling with molecular parameters. The present quantitative success lies in the allowance of the *in vivo* and *in vitro* differences, and of various noise contributions. However, those four quantities may be directly determined biologically.

There are three different types of experimental data to determine the dynamical elements of the local potential function for a genetic switch, the degradation (the friction in physical sciences), and the barrier height (Fig. 7). The first type is the protein distribution around each of the epigenetic states. It is given mathematically by

$$\rho(\mathbf{N}) = \rho_0 \exp(-U(\mathbf{N})), \qquad \text{Eq.(18)}$$

here $\rho_0$ is a normalization constant. The protein distribution is explicitly measurable. Once $\rho(\mathbf{N})$ is measured experimentally, the local potential $U(\mathbf{N})$ near the potential minima can be determined as: $U(\mathbf{N}) = -\ln(\rho(\mathbf{N})) + \ln(\rho_0)$.

The second type of experimental data is the relaxation time, a measure of how long it would take for the system to return to its local equilibrium after a small perturbation. It is determined by both the potential near the minimum and the degradation,

$$\tau = \eta/U'', \qquad \text{Eq.(19)}$$

here η is the strength of friction, which gives the friction matrix Λ along the path of relaxation. U″ the second derivative of potential. Since potential can be obtained from ρ(**N**), relaxation time can be used to obtain friction: η = τ U″.

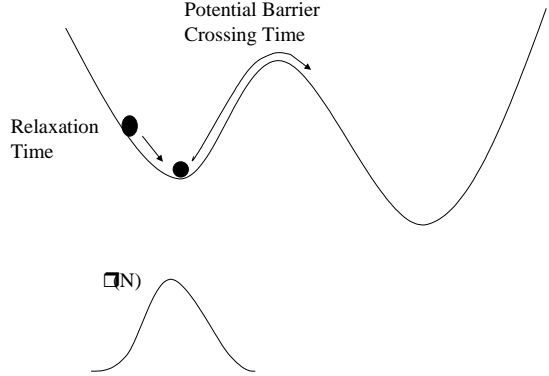

Figure 7. Three types of experiments directly probe the dynamical structure of a genetic switch and determine the dynamic components. The measurement of protein number distribution determines the potential function at each of the epigenetic state. The additional information on the relaxation time determines the strength of friction. The lifetime of each epigenetic state determines the height and shape of potential barrier.

The third type of experimental data is the lifetime of epigenetic states, the measure of switching rate from one state to another. The probability of phage evolving from one epigenetic state of growth to another is given by the Kramers rate formulae, our Eq.(16),

$P = \omega_0 \exp(-\Delta U_b)$ ,

where $\Delta U_b$ is barrier height and $\omega_0$ is the attempt frequency. $\omega_0$ is given by the friction and the curvature of the potential barrier. The curvature of the potential is related to the height of the potential barrier and the shape of potential near its minimum. Therefore $\Delta U_b$ can be determined from the lifetime of its epigenetic state: $\Delta U_b = \ln(\omega_0) - \ln(P)$.

The genetic switch for phage λ is a complex dynamical system. It took decades of ingenious experimental research and laborious work to collect parameters needed for this mathematical modeling. For a more complicated system, resource and time may limit the ability to study each molecular element in such a great detail. A method that is less demanding on the details yet still can capture the main features is of great interest. Above dynamical structure analysis provides such a guidance to build such a phenomenological model, as illustrated in .

We emphasize that the quantities introduced in dynamical structure theory, the friction, the potential gradient, the transverse force and the stochastic force associating with the friction are all measurable quantities at the given description level. These are quantities similar to temperature, pressure, and free energy in thermodynamics, which can be determined by the microscopic details but can also be measured independent of those details. Once the relationship between these quantities are established, as shown in Eq.(10), we are ready to write down the effective equation of motion for the network without resorting to details.

### 5.6. Stochasticity, robustness, cooperation, and efficiency

Starting from realization that noise is important in the initiation of transcription in phage $\lambda$ [58], stochasticity has been increasingly viewed as one of the most important elements in the dynamical modeling of biological processes [106-108]. Both the intrinsic and extrinsic noises are shown to exist in biological processes and are analyzed theoretically [90-93]. The parameterization in the form of Eq.(17) is a way to account for both noise contributions.

Robustness has been viewed as one of the central features in biological processes [109-111]. Numerous recent studies have established its importance [112-117]. Combing with stochasticity, the work reviewed here [9,64] established a quantitative criterion, Eq.(16), for the robustness. The potential landscape function, $U(\mathbf{N})$ in Eq.(10), emerged from the stochastic dynamics provides a graphic representation for the robustness. Those results again confirm the importance of noise.

Finally, we wish to point out that a complete rigid system, that is, an absolute robust system with no flexibility, is not viable from the evolution point of view [118,119]. Such a structure would not survive the stringent evolutionary process. This may be illustrated in the switch efficiency discussed above. It is also implied by the ability that the phage $\lambda$ can switch from lysogenic to lytic state when provoked [8], a feature can be captured based on the present minimum model, though no explicit and detailed mathematical analysis has been published yet along this direction. Thus, the stochasticity seems to provide the critical link to understand both robustness and flexibility. We believe such a feature can indeed be understood from the evolutionary point of view [119].

## 6. Perspective on Mathematical Modeling

### 6.1. Major prediction of the minimum quantitative modeling

We have shown that thanks to continuous theoretical and experimental efforts the minimum quantitative modeling has achieved the status of quantitative agreement with experimental biological data. New predictions, such as protein distributions, are made and discussed in Ref.[9]. There is one prediction on the cooperative energy which stands out as an excellent indicator for success of both recent theoretical and experimental efforts.

Ever since 1980's, it has been found that it is rather difficult to model stability of $\lambda$ switch with known parameter constraints [48,57], even allowing the possibility of up to 30% difference between *in vivo* and *in vitro* parameter values [51]. It has been found the cooperation energy would play an important role [120,121]. Thus, it has been hypothesized that additional effects beyond the minimum model would be needed. One of the most promising one is a stronger cooperative effect. Indeed, independent of theoretical need, additional effect, the looping, has been found experimentally [89,97].

A brief account of this effort may be relevant. Four years ago four of the present authors began a mathematical study on phage λ. Nevertheless, tried all known methods to us at that time we could not solve the stability puzzle. Effectively we were in the same situation as what reported in Ref's.[57,67]. One of major problems for us was that even allowing the possibility to vary the parameter values drastically in the name of *in vivo* and *in vitro* difference, the parameter space appears too big for an effective research. This difficulty was partially verified in retrospective in Ref.[51], where the change of parameters value appeared not large enough in order to explain the experimental within the minimum quantitative model. Out of this frustration it was realized three years ago that one must have an effective quantification criterion for the stability. It turned out that the landscape idea rooted deeply in both physics and biology appears to be such a candidate. Driving by this need for quantification, a mathematically consistent construction method for such a landscape function was quickly discovered. With this new method it was relatively easy to explore bigger parameter space. One critical parameter, the cooperative energy, was then found to double its value in order to have the desired stability, the value in bold face value for ΔG(cooperative) of -3.7 kcal/mol in Table II. We note this value is about twice of what tried in Ref.[51].

Interestingly such a value was indeed observed in an independent biological experiment [89]. In writing this review, we further noticed that such a big value was suggested to be possible in an independent theoretical investigation based on thermodynamic consideration [88]. Given all those independent efforts: theoretical [9,88] and experimental [89,97], successful [64] and failed [51], present authors believe that such an agreement between the theoretical prediction [64] and experimental value [89] on the cooperative energy may not be accidental. It indicates that that the minimum quantitative model may have indeed captured the essential biological features of this genetic switch, with its first nontrivial and verified prediction.

**6.2. Relation to other modeling methodologies**

There has been a tremendous amount of literature on biomodeling and biocomputation. It is both impossible in the present article to give an adequate survey of various methodologies. Nevertheless we would like to present the following two classification schemes, according to mathematical and to scientific structures, to place our method in a broader context.

From a mathematical point of view, a modeling may be classified according to whether it is discrete or continuous and whether it is deterministic or stochastic. The classical modeling of deterministic and discrete is the Boolean logic circuit [122]. The works of Shea and Ackers [48] as well as others [57,123,124] are the fine examples of deterministic and continuous modeling. Examples of stochastic and continuous modeling are Arkin *et al*. [58], Aurell and Sneppen [59]. Naturally, there are methods of combining various features. One of such fine examples is the hybrid of continuous and discrete modeling of Tchuraev and Galimzyanov [125]. Of those modeling methods according this classification, the simplest one is that based on the Boolean logic circuit. It is clearly an approximation, but in many cases well serves specific biological purposes. It is in fact currently the dominant modeling and presentation methodology in biology. The most difficult but most detailed modeling is the continuous and stochastic formulation. Many of its predictions are necessarily probabilistic in nature, corresponding nicely to biological phenomena. Our present method belongs to this last category. Nevertheless, we should point out that no method would be definitely better than the rest. The choice of modeling method must be appropriate to biological questions under addressed.

From the scientific structure point of view, the modeling may be classified into the first principle modeling and the completely empirical modeling. It is believed that chemical reactions and other physical processes lie behind various biological processes. Hence it should be possible to predict biological functions based on the physical-chemical principles. This first principle modeling methodology has been followed by Shea and Ackers [48] and by Reinitz and Vaisnys [57], by many others [58,59,123]. Ours is also of this type. The advantages of first principle modeling are that it shows the unity of sciences and that additional insight and information can be obtained from the lower level scientific descriptions. The evident disadvantage of this modeling, in addition to the difficulty of specifying all needed micro-parameters, is that higher level processes often show emerging phenomena. It is difficult to make predictions based on the properties of system's components. The outstanding stability puzzle of phage $\lambda$ genetic switch [51,59], one of simplest possible living genetic switch, and the enormous effort [9,48,51,57-59,67] to quantitatively model its behavior, clearly illustrate this situation.

The other extreme, as comparing to the first principle modeling, is to treat the system in question autonomously, inferring its properties completely from empirical studies. Statistical analysis methods play a dominant role in this approach. The advantage of this modeling is that it establishes the independent role of the investigating scientific layer. It is consistent with the view that at each scientific level autonomous laws can be uncovered. Equipped with biological insights, it has been employed successfully in numerous biological studies. The proposals of Waddington [126] on developmental landscape and of Monod and Jacob [127] on gene regulation mechanism are such fine examples. In reality, particularly in molecular biology, what typically encountered is in between those two extremes, as demonstrated by the Boolean logic circuit modeling[5] [122] and by others [123-125,128]. Interestingly, even in the empirical modeling setting, we have, however, briefly discussed the direct and transparent connection of our method to empirical data in subsection 5.5., though it is rooted in first principle modeling. This suggests that our method can be used in this extreme, too. Further investigation in this direction should be carried out.

The statistical analysis also suggests an important issue in mathematical modeling: the number of variables and the associated problem of "curse of dimension". Regarding to the matching of data to appropriate modeling methodologies, there is another issue of "parsimony of experimental data", which is particularly acute in a real time modeling at present. We will not discuss those issues here.

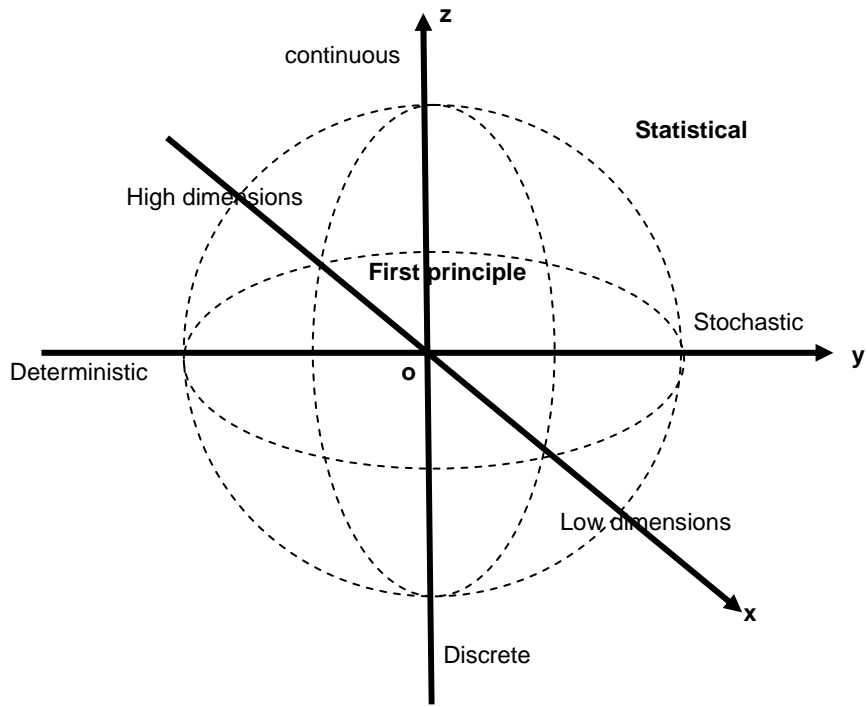

Figure 8. The schematic diagram of modeling methodologies. The coordinates are classified according to mathematical descriptions which define a three dimensional space: many degrees of freedom (high dimensions) to few degrees of freedom (low dimensions) (x-direction); from deterministic to stacohastic (y-direction); discrete to continuous (z-direction). The broken lines which defines a sphere are classified according to scientific understandings: from first principles (inside) to statistical methods (outside). Both scientific methodologies encompass all mathematical descriptions.

To summarize the unique characters of the present novel method, the stochastic dynamical structure analysis, mathematically it is a continuous and stochastic formulation with four dynamical elements. In terms of scientific structure, it is equivalent to the first principle modeling. Nevertheless it can be directly related to empirical data to establish its autonomy.

**6.3. Literature sampling**

First of all, the book by Ptashne [8] on phage $\lambda$ is a must read, where an excellent review on experimental work up to 2004 can be found. Because both the $\lambda$ repressor and the *lac* operon are instrumental in the current molecular and synthetic biological studies, the book by Mulller-Hill [129] is another must read. A good summary of earlier study on phage lambda can be found in Ref.[130]. A broader and recent general review can be found in Ref.[131].

For a recent phage λ studies reviewed from the switch point of view, Ref.[132] is a good start. Five switches were identified there for the developmental process. The stability puzzle was put into sharp focus in Ref.[67]. Looping study has been studied in Ref.[88] theoretically and in Ref.[89,97] experimentally. The phage from evolutionary point of view was studied in Ref.[133]. More studies on the role of Cro can be found in Ref.[84] and [85]. More interesting dynamical behaviors were reported experimentally in Ref.[134] and [135]. The effect of degradation time on stability was considered theoretically in Ref.[136]. Various other features on genetic switch have been recently explored in Ref's.[137-142].

## 7. Third Age of Phage

Because of its enormous biomass in the biosphere on Earth, the importance of phage has already been recognized in the current ecological study [47,143,144], and title of this section is borrowed from Ref.[144]. The presentation of theoretical effort in this chapter has shown that the phage has been playing an important role in the study of fundamental biology in the post Human Genomic era, too, along with the experimental effort [145].

The quantitative and detailed modeling of gene regulatory networks is evidently at its beginning. There are numerical possibilities to go beyond the minimal quantitative modeling reviewed here. For example, even for the phage lambda genetic switch, it is a simplification. It would be desirable to have quantitative demonstration on how some of the in vivo and in vitro differences arise by incorporating more degrees of freedoms, such as the left operon and what would be the quantitative differences. A further extension would be to model all five switches of the lambda developmental processes [132], to obtain a comprehensive quantitative understanding of the whole process. Deep biological questions, such as why the phage chooses such a structure or what are the evolution principles guided this choice [66,128], not discussed yet adequately by any measure. Nevertheless, we do wish to point out that the study of phage λ genetic switch has revealed a novel mathematical structure [9] and already put one of the deep rooted concepts in biology, the landscape [126,146-148], back on a firm mathematical and biological ground. Numerical recently quantitative phage studies [9,49-51,58,59,66,67,74-76,132-142] have pushed our understanding of systems biology onto another level, complementary to high-throughput and large-scale analyses. It should be optimistic that its study will generate more new biological understandings and will have an influence beyond biology, as biology has already inspired the theories of general systems [149] and cybernetics [150]. In view of its past glorious successes (For example, one may count how many Nobel Prize winners in physiology and medicine and in chemistry appeared in http://www.asm.org/division/m/blurbs/secrets.html#top.), we are confident that more discoveries are waiting ahead.

## Acknowledgements

We thank J.J. Collins, J.W. Little, B. Muller-Hill, S.M. Stoylar, G. Wegrzyn for critical comments and M. Mossing for an updating on literature. This work was supported in part by the Institute for Systems Biology (L.H. and D.G.) and by a USA NIH grant under HG002894 (P.A.).